\documentclass[10pt]{article}
\usepackage{amsmath}
\usepackage{amssymb}
\usepackage{amsfonts}
\newcommand{\ba}{\begin{array}{l}}
\newcommand{\ea}{\end{array}}
\newcommand{\beq}{\begin{equation}}
\newcommand{\eeq}{\end{equation}}
\newcommand{\bea}{\begin{eqnarray}}
\newcommand{\eea}{\end{eqnarray}}
\usepackage{epsfig}
\usepackage{graphicx}
\usepackage{color}

\definecolor{dyellow}{rgb}{1.,0.8,.0}
\definecolor{myblue}{rgb}{.1,.1,.7}
\definecolor{dcyan}{rgb}{.0,.6,.6}
\definecolor{dmagenta}{rgb}{0.6,0.0,0.6}
\definecolor{brown}{rgb}{0.6,0.2,0.}
\definecolor{darkblue}{rgb}{.0,.0,0.5}
\definecolor{darkred}{rgb}{0.75,0.0,0.0}
\definecolor{orange}{rgb}{1.,.6,.0}
\definecolor{dorange}{rgb}{0.8,.4,.0}
\definecolor{darkgreen}{rgb}{0.0,0.6,0.0}
\definecolor{purple}{rgb}{.4,.0,.4}

\def\la{\lambda}

\def\bc{\begin{center}}
\def\ec{\end{center}}
\def\be{\begin{eqnarray}}
\def\ee{\end{eqnarray}}

\newcommand{\omits}[1]{}

\begin{document}
\baselineskip=14pt
%

%
%

\begin{center}
{\Large \bf { Towards a CPT Invariant Quantum Field Theory\\
 on Elliptic de Sitter Space }}\\
  \vspace*{1cm}
Zhe Chang \footnote{changz@mail.ihep.ac.cn},  Xin
Li \footnote{lixin@mail.ihep.ac.cn}\\
\vspace*{0.2cm} {\sl Institute of High Energy Physics,
Chinese Academy of Sciences\\
P. O. Box 918(4), 100049 Beijing, China}\\

\bigskip

\end{center}
\vspace*{2.5cm}

%
\begin{abstract}
Consequences of Schr\"{o}dinger's antipodal identification on
quantum field theory in de Sitter space are investigated. The
elliptic $\mathbb{Z}_2$ identification provides observers with
complete information.  We show that a suitable confinement on
dimension of the elliptic de Sitter space guarantees the existence
of globally defined spinors and orientable $dS/\mathbb{Z}_2$
manifold. In Beltrami coordinates, we give exact solutions of scalar
and spinor fields. The CPT invariance of quantum field theory on the
elliptic de Sitter space is presented explicitly. \vspace{2cm}
\begin{flushleft}
\end{flushleft}
\end{abstract}

\newpage
%
%
\section{Introduction}
Quantum field theory in curved space is of fundamental interest in
the understanding of a conceptual unification of general relativity
and quantum mechanics. Quantum field theory on de Sitter space is of
great interest for three reasons. First, it provides a promising
``inflationary" model of the very early universe \cite{1}. The
second, it is a highly symmetric curved space, in which one can
quantize fields and obtain simple exact solutions of fields.
Finally, the astronomical observations on supernova and the cosmic
microwave radiation suggest that the real universe resembles de
Sitter space \cite{2}.

The de Sitter space contains the past and future null infinity
${\cal I}^+$ and ${\cal I}^-$. They are the surfaces where all null
geodesics originated and terminated.  This fact rises absurdities in
the study of quantum field theory on de Sitter space. The first one
is that a single immortal observer in de Sitter space can see at
most half of the space. The second problem, as pointed in \cite{3},
is that if enough matter is presented, the asymptotic de Sitter
space may collapse at ${\cal I}^-$  in finite time. And there will
be no future asymptotic de Sitter region ${\cal I}^+$ at all. Even
if the collapse does not occur, presence of matter alters the causal
structure also\cite{4}. The third and most fatal problem arises in
trying to define the analogue of an S matrix. In quantum field
theory, asymptotic incoming and outgoing states are properly defined
only in the asymptotic regions of spacetime. However, an observer
can not denote the states continuously from past infinity to future
infinity. Consequently, the matrix elements of S-like matrix are not
measurable. When one considers quantum gravity in asymptotically de
Sitter space,  the situation becomes even more serious.  As has been
pointed out by Witten\cite{5},  the only available pairing between
incoming states and outgoing states,  CPT, should be used to obtain
an inner product in the Hilbert space. The conventional formulation
of quantum field theory and string theory is based on the existence
of an $S$ matrix.

In order to construct a consistent quantum field theory in de Sitter
space, one should deal with the problems shown above carefully. The
elliptic interpretation pioneered by Schr\"odinger\cite{6} has been
gaining interests again. In the elliptic de Sitter space, each point
is identified with its antipode. This simple $\mathbb{Z}_2$
identification makes all observers able of gaining complete
information about all events. For black holes, the identification
destroys the thermal features and  Fock space construction\cite{7,
8}.

Quantum field theory on the elliptic de Sitter space has been
discussed extensively[9--11]. Unfortunately, a general
$d$-dimensional elliptic de Sitter space $dS_d/\mathbb{Z}_2$ is
non-orientable. This is a fatal problem for constructing a quantum
field theory. In this paper, we have calculated the first
Stiefel-Whitney class of the elliptic de Sitter space and showed
that the manifold $dS_d/\mathbb{Z}_2$ is orientable for odd $d$.
Furthermore, we have noted that the second Stiefel-Whitney class
$\omega_2(dS_d/\mathbb{Z}_2)$ vanished while the dimension of
elliptic de Sitter space equals to $3~{\rm mod}~4$. Thus, for
$(3+4n)$-dimensional elliptic de Sitter space, there exists globally
defined spinor fields. In Beltrami coordinates, we have obtain exact
solutions of scalar field and spinor field. The CPT invariance of
the quantum field theory on the elliptic de Sitter space is shown
explicitly.

This paper is organized as follows. In Sec.2, we briefly review
properties of de Sitter space and present arguments on why we should
introduce the elliptic identification. In Sec.3, we investigate
geometrical properties of the elliptic de Sitter space and present
explicitly a confinement on dimension of the elliptic de Sitter
space. This confinement guarantees the manifold $dS_d/\mathbb{Z}_2$
orientable and existence of globally defined spinor field. Sec.4 is
devoted to explicit construction of scalar and spinor field in
Beltrami coordinates on the elliptic de Sitter space. In Sec.5,
field quantization in the elliptic de Sitter space is discussed in a
general way.  The $CPT$ invariance of the scalar quantum field
theory is shown. In the final, we give conclusions and remarks in
Sec.6.

\section{ Elliptic identification}
De Sitter space is a maximally symmetric space\cite{b1,b2}. In the
$(d+1)$-dimensional Minkowski space with metric $\eta_{ab}={\rm
diag}(-1, \underbrace{1, \cdots, 1}_{d})$, the de Sitter space
$dS_d$ is a hyperboloid with embedding equation \be
\eta_{ab}X^aY^b=a^2. \ee Here the constant  $a$ is a parameter with
units of length called the de Sitter radius.  In the paper, we use
the lower case letter $x$ denotes a $d$-dimensional coordinate on
$dS_d$ and the upper case letter $X$ denotes the corresponding
$(d+1)$-dimensional coordinate in the embedding space.

The de Sitter space has a constant positive curvature
$R=d(d-1)a^{-2}$. Its symmetry group is $O(1, d)$, which consists of
four disconnected components $G_I, G_T, G_S$ and $G_{ST}$. Here $I$,
$T$ and $S$ denote the identity, time reversal and space reflecting,
respectively, \be I\equiv {\rm diag}(1, \underbrace{1, \cdots,
1}_{d})~,~~~~ T\equiv {\rm diag}(-1,\underbrace{1, \cdots,
1}_{d})~,~~~~~ S\equiv {\rm diag}(1, \underbrace{-1, \cdots,
-1}_{d}). \ee $G_I$ denotes the component which contains the
identity. $G_T$ is obtained by acting $T$ on $G_I$,  \be G_T\equiv
\{g\cdot T|g\in G_I\}. \ee In a simlar way, we can get $G_S$ and
$G_{ST}$ components. Define an antipodal transformation $A$, which
is an element of $G_{ST}$, as \be A\equiv {\rm diag}(-1,
\underbrace{-1, \cdots, -1}_{d}). \ee The antipodal transformation
sends a point $p$  with coordinate $X^a(p)$ to its antipode
$\bar{p}$  with coordinate $X^a(\bar{p})[=-X^a(p)]$. Since $A$ is an
element of $G_{ST}$, it reverses direction of time.

It is convenient to introduce a quadratic real form \be Z(x,
y)=a^{-2}\eta_{ab}X^a(x)Y^a(y). \ee Then, the geodesic distance
between points $p(x)$ and $q(y)$ can be written into the form \be
d(x, y)=a\arccos Z(x, y). \ee From the definition of the quadratic
real form $Z(x, y)$, we know that it changes sign when one point
$p(x)$ is sent to its antipode $\bar{p}(\bar{x})$, \be \label{Z}Z(x,
y)=-Z(\bar{x}, y). \ee

To investigate properties of the elliptic de Sitter space, one may
begin with discussing on relations between point and its antipode in
ordinary de Sitter space\cite{9}. First, the point $p$ and its
antipode $\bar{p}$ is always space-like separated, since
$Z(p,\bar{p})=-a^2X^2(p)<0$. Moreover, the interiors of the light
cones of the point $p$ and its antipode $\bar{p}$ do not intersect.
If such intersection exist and a point $q(Y)$ is belong to the
intersection, one would have\cite{10}
 \be (Y+X)^2=(Y-X)^2=0 \Rightarrow
Y^2=-a^2. \ee  So the point $q(Y)$ does not lie on the de Sitter
hypersurface. This shows that the light cones of two antipodal
points within de Sitter space do not intersect. All the events
observed by a single immortal observer are occurred at points inside
the future event horizon which cover only one half of de Sitter
space. The other half made up of antipodal points of the points
inside the future event horizon. From the above discussion, we know
that there is no event horizon and no closed timelike curves in the
elliptic de Sitter space.

It should be noticed that the elliptic identification maps ${\cal
I}^+$ and ${\cal I}^-$ to each other.  This will be useful for
constructing an $S$ matrix.

However, the quotient space $dS/\mathbb{Z}_2$ is
non-simply-connected and non-time-orientable for arbitary dimension.
The operator $A$ reverses direction of time. Non-time-orientable
elliptic de Sitter space may rise causality problems\cite{10}.

In the rest of the paper, we set $a=1$.

\section{Geometrical properties}
The general elliptic manifold $dS_d/\mathbb{Z}_2$ is non-orientable.
For the non-orientable manifold, we can not use the ``Stokes
formula". Fortunately, we will show that the manifold
$dS_d/\mathbb{Z}_2$ is orientable while $d$ is odd.

A real vector bundle $E$ over $M$ is called orientable if the
structure group of $E$ can be reduced to $GL^+(\mathbb{R}, n)$. A
manifold is called orientable if its tangent bundle is orientable.
There is a theorem, which indicates that the bundle $E$ with its
first Stiefel-Whitney class $\omega_1(E)=0$ must be
oriented\cite{MGM3}.

Manifolds with a non-vanishing second Stiefel-Whitney class do not
admit globally defined spinors\cite{14}. For constructing a quantum
spinor field theory in elliptic de Sitter space globally, we need
the elliptic de Sitter space to have  a vanishing second
Stiefel-Whitney class [$\omega_2(E)=0]$.

In order to calculate the first and second Stiefel-Whitney class of
the elliptic de Sitter space, we introduce a line bundle over
$\mathbb{R}\mathbf{P}^n$---$E(\gamma^n)$,
 \be E(\gamma^n)=\{(\{\pm
x\}, v)\in \mathbb{R}\mathbf{P}^n\times\mathbb{R}^{n+1}\mid
v=\lambda x, \lambda\in\mathbb{R}\}\nonumber,\\ \pi:
E(\gamma^n)\rightarrow \mathbb{R}\mathbf{P}^n: (\{\pm x\}, v)\mapsto
\{\pm x\}. \ee

There are two useful theorems on $\gamma^n$\cite{16} for calculating
the Stiefel-Whitney classes of the elliptic de Sitter space.\\
Theorem 1: {\rm $\omega_1(\gamma^n)$ generates
$H^1(\mathbb{R}\mathbf{P}^n,\mathbb{Z}_2)$, thus $\gamma^n$ is
non-orientable}.\\
Theorem 2: {\rm The Whitney sum
$T(\mathbb{R}\mathbf{P}^n)\oplus\varepsilon^1$ is isomorphic to the
(n+1)-fold Whitney sum
$\underbrace{\gamma^n\oplus\gamma^n\cdots\oplus\gamma^n}_{n+1}$}
Here $T(\mathbb{R}\mathbf{P}^n)$ is the tangent bundle of
$\mathbb{R}\mathbf{P}^n$ and $\epsilon^1$ is a trivial line bundle
over $\mathbb{R}\mathbf{P}^n$.\\ From the mapping $i:
S^n/\mathbb{Z}_2\rightarrow(\mathbb{R}^1\times S^n)/\mathbb{Z}_2$ (a
homotopy equivalence), we get the induced map \be\label{whitney}
i^\ast T(\mathbb{R}^1\times S^n)/\mathbb{Z}_2&=&(\mathbb{R}^1\times
S^n\oplus T(S^n))/\mathbb{Z}_2\nonumber\\ &=&\gamma^n\oplus
T(\mathbb{R}\mathbf{P}^n). \ee By making use of relation
(\ref{whitney}) and axioms of Stiefel-Whitney class, we get\cite{pc}
\be\label{whitney eq} \omega(T(\mathbb{R}^1\times S^{d-1}\oplus
T(S^{d-1}))/\mathbb{Z}_2)&=&\omega((\mathbb{R}^1\times S^{d-1}\oplus
T(S^{d-1}))/\mathbb{Z}_2)
\nonumber\\&=&\omega(\gamma^{d-1})\smallsmile\omega(T(\mathbb{R}\mathbf{P}^{d-1}))
\nonumber\\&=&\omega(\gamma^{d-1})\smallsmile\omega(T(\mathbb{R}\mathbf{P}^{d-1})+\varepsilon^1)
\nonumber\\&=&\omega(\gamma^{d-1})\smallsmile\omega(d\gamma^{d-1})
\nonumber\\&=&(\omega(\gamma^{d-1}))^{d+1}=(1+\omega_1(\gamma^{d-1}))^{d+1}
\nonumber\\&=&1+(d+1)\omega_1(\gamma^n)+\frac{(d+1)d}{2}\omega_1(\gamma^n)^2+\cdots
.\ee From the second and third part of the above equations, we
obtain\be \omega_1(i^\ast T(\mathbb{R}^1\times
S^{d-1})/\mathbb{Z}_2))&=&(d+1)\omega_1(\gamma^n)=0,~~~{\rm
for}~~~d=1~~{\rm mod}~~2 ~~ \nonumber\\
\omega_2(i^\ast T(\mathbb{R}^1\times
S^{d-1})/\mathbb{Z}_2))&=&\frac{(d+1)d}{2}\omega_2(\gamma^n)=0,~~~{\rm
for}~~~d=3~~{\rm mod}~~4.\ee Since $i$ is a homotopy equivalence,
one has \be \omega_j(i^\ast TM)=0\Leftrightarrow\omega_j(TM)=0. \ee
Thus, we can conclude that the $(3+4n)$-dimensional elliptic de
Sitter space is orientable and there exists globally defined spinors
on it.

\section{ Beltrami coordinates and Fock space}
\subsection{Manifold structure of $\mathbb{R}\mathbf{P}^d$  and Beltrami coordinates of
elliptic de Sitter space }
Before discussing quantum field theory on
elliptic de Sitter space in Beltrami coordinates, we recall the
nature manifold structure of the projective spaces
$\mathbb{R}\mathbf{P}^d$\cite{13}.  Consider the set of all straight
lines in $R^{d+1}$ passing through the origin. Since such a straight
line is completely determined by any direction vector, and since any
non-zero scalar multiple of any particular direction vector serves
equally well, we may take these straight lines as the points of
$\mathbb{R}\mathbf{P}^d$. Now each of these straight lines
intersects the sphere $S^d$ (with equation
$(X^0)^2+\cdots+(X^d)^2=1$) at exactly two (diametrically opposite)
points. Thus the points of $\mathbb{R}\mathbf{P}^d$ are in
one-to-one correspondence with the pairs of diametrically opposite
points of $d$-sphere. We may therefore think of projective space
$\mathbb{R}\mathbf{P}^d$ as obtain from $S^d$ by ``glueing", as they
say, that is by identifying, diametrically opposite points.
$\mathbb{R}\mathbf{P}^d$ is topologically equivalent to
$S^d/\mathbb{Z}_2$. This will be useful for us to discuss elliptic
de Sitter space.

To give a manifold structure on the projective spaces
$\mathbb{R}\mathbf{P}^d$, let us return to original characterization
of $\mathbb{R}\mathbf{P}^d$ as consisting of equivalence classes of
non-zero vectors in the space $\mathbb{R}^{d+1}$ with coordinates
$X^0, \cdots, X^d$. Let $U_q$ ($q=0, 1, \cdots, d$) denote the set
of equivalence classes of vectors ($X^0, \cdots, X^d$) with
$X^q\neq0$. On each such region $U_q$ of $\mathbb{R}\mathbf{P}^d$ we
introduce the local coordinates \be x_q^1=\frac{X^0}{X^q}, \cdots,
x_q^q=\frac{X^{q-1}}{X^q}, x_q^{q+1}=\frac{X^{q+1}}{X^q}, \cdots,
x_q^d=\frac{X^d}{X^q}. \ee Clearly the regions $U_q$ ($q=0, 1,
\cdots, d$) cover the whole of projective $d$-dimensional space. The
general formulae for the transition functions on $U_j\bigcap U_k$
can be obtained from those for $U_0\bigcap U_1$ by the appropriate
replacement of indices. In the region $U_0\bigcap U_1$, where both
$X^0, X^1\neq0$, the transition function from $(x_0)$ to $(x_1)$ is
obviously \be x_1^1=\frac{1}{x_0^1}, x_1^2=\frac{x_0^2}{x_0^1},
x_1^3=\frac{x_0^3}{x_0^1}, \cdots, x_1^d=\frac{x_0^d}{x_0^1}. \ee

The Beltrami coordinates in elliptic de Sitter space can be defined
in the same way as in $\mathbb{R}\mathbf{P}^d$, \be
x^a=\frac{X^a}{X^d},~~~~ (a=0, \cdots, d-1),~~~~X^d\neq0. \ee Here
we simply choose the $U_d$ patch of the elliptic de Sitter space to
discuss. On the other patches, same discuss can be given. For the
sake of convenience, we define \be \sigma(x^a, x^b)\equiv1+\sum_{a,
b=0}^{d-1}g_{ab}x^ax^b, \ee where $g={\rm diag}(-1, \underbrace{1,
\cdots, 1}_{d-1})$. The metric of elliptic de Sitter space in the
Beltrami coordinates is given as\cite{c}\cite{24} \be
\mathbf{d}s^2=\label{beltrami metric}-\frac{\mathbf{d}x
g(I-x^Txg)^{-1}\mathbf{d}x^T}{1-xgx^T}. \ee In order to compare with
the global time which defined in the global coordinates $(t,
\Omega)$, we write the metric (\ref{beltrami metric}) in terms of
coordinates $(X^0, x^1, \cdots, x^{d-1})$.  The relation between
$X^0$ and the global time $\tau$ is $X^0=\sinh\tau$. By making use
of the relations between the coordinates $(x^a)$ and $(X^0, x^1,
\cdots, x^{d-1})$ \be \sigma(x^a,
x^b)&=&\frac{1+\vec{x}\vec{x}^T}{1+X^0X^0},
\\x^0x^0&=&\frac{X^0X^0}{1+X^0X^0}(1+\vec{x}\vec{x}^T), \ee we deform the
metric (\ref{beltrami metric}) as \be \label{first metric}
\mathbf{d}s^2=-\frac{\mathbf{d}X^0\mathbf{d}X^0}{1+X^0X^0}+(1+X^0X^0)\frac{\mathbf{d}\vec{x}(I+\vec{x}^T\vec{x})^{-1}\mathbf{d}\vec{x}^T}{1+\vec{x}\vec{x}T},
\ee where $\vec{x}$ means just take the spatial part. Writing the
spatial section of the metric (\ref{first metric}) in spherical
coordinates, we can get \be \label{second metric}
\mathbf{d}s^2=-\frac{\mathbf{d}X^0\mathbf{d}X^0}{1+X^0X^0}+(1+X^0X^0)[(1+\rho^2)^{-2}\mathbf{d}\rho^2+(1+\rho^2)^{-1}\rho^2\mathbf{d}\Omega^2_{d-2}].
 \ee
Rewrite metric (\ref{second metric}) in terms of Beltrami time
$\tau$ as \be
\mathbf{d}s^2=-\mathbf{d}\tau^2+\cosh^2\tau[(1+\rho^2)^{-2}\mathbf{d}\rho^2+(1+\rho^2)^{-1}\rho^2\mathbf{d}\Omega^2_{d-2}].
\ee

\subsection{Scalar field }
Let us consider a scalar field in $dS_d/\mathbb{Z}_2$ with action

\begin{equation}
S=-\frac{1}{2}\int
d^dx\sqrt{g}\left[\left(\nabla\phi\right)^2+m^2\phi^2\right]~.
\end{equation}
By writing the scalar field $\phi$ into variable-separating form \be
\phi(\tau, \rho, \Omega)=T(\tau)U(\rho)Y_{l\mathbf{m}}(\Omega), \ee
we can transform the equations of motion of the scalar field in
$dS_d/\mathbb{Z}_2$ as\be
\label{time}\cosh^2\tau\ddot{T}(\tau)+(d-1)\sinh\tau\cosh\tau\dot{T}(\tau)+(m^2\cosh^2\tau-\epsilon)T(\tau)&=&0, \\
\label{radi}U''(\rho)+\frac{d-2+2\rho^2}{\rho(1+\rho^2)}U'(\rho)-\left[\frac{\epsilon}{(1+\rho^2)^2}+\frac{l(l+d-3)}{\rho^2(1+\rho^2)}\right]U(\rho)&=&0,\\
\vspace{2mm}
 [\Delta_{S^{d-2}}+l(l+d-3)]Y_{l\mathbf{m}}(\Omega)&=&0, \ee where
 $\epsilon$ is an arbitrary real constant.
Here the $Y_{l\mathbf{m}}$ is not in standard form. In terms of
usual spherical harmonics $U_{l\mathbf{m}}$, the $Y_{l\mathbf{m}}$
is expressed as \be
Y_{l\mathbf{m}}=\frac{1}{\sqrt{2}}[e^{i\pi/4}U_{l\mathbf{m}}+e^{-i\pi/4}U_{l\mathbf{m}}^\ast].
\ee The spherical harmonic functions $Y_{l\mathbf{m}}$ are also
orthnormal and form a complete set. In particular, we have
\be\label{antipodal Ylm}
Y_{l\mathbf{m}}(\Omega_A)=(-1)^lY_{l\mathbf{m}}^\ast(\Omega). \ee

To solve the equation (\ref{time}), first we change its variable as
$X^0$. Then, the equation (\ref{time}) transforms
as\be\label{equation Tx}
(1+{X^0}^2)^2\frac{\mathbf{d}^2T(X^0)}{\mathbf{d}{X^0}^2}+d\cdot
X^0(1+{X^0}^2)\frac{\mathbf{d}T(X^0)}{\mathbf{d}X^0}+\left[m^2(1+{X^0}^2)-\epsilon
\right]T(X^0)=0. \ee By substituting \be
(1+{X^0}^2)^{\frac{d-2}{4}}T(X^0)=f(X^0), \ee we get \be\label{X0}
(1+{X^0}^2)\frac{\mathbf{d}^2f(X^0)}{\mathbf{d}{X^0}^2}+2X^0\frac{\mathbf{d}f(X^0)}{\mathbf{d}X^0}+\left[m^2-\frac{d(d-2)}{4}+\frac{(d-2)^2-4\epsilon
}{4(1+{X^0}^2)}\right]f(X^0)=0. \ee The solution of Eq. (\ref{X0})
is \be f(-iX^0)=\left\{
                   \begin{array}{l}
                   P_{\nu}^{\mu}(-iX^0) \\
                   \newline              \\
                   Q_{\nu}^{\mu}(-iX^0)
                   \end{array}\right.  ~, \ee
where\be
\nu(\nu+1)=\frac{d}{2}\left(\frac{d}{2}-1\right)-m^2,~~\mu=\sqrt{\left(\frac{d}{2}-1\right)^2-\epsilon
}~, \ee and $P_{\nu}^{\mu}(-iX^0)$,~~ $Q_{\nu}^{\mu}(-iX^0)$ are the
first and second kind of associate Legendre function respectively.

By substituting \be
U(\rho)=\rho^l(1+\rho^2)^{\frac{\kappa}{2}}F(\rho), \ee where\be
\kappa^2-\kappa(d-2)=-\epsilon,  \ee the equation (\ref{radi})
transforms as \be
(1+\rho^2)\frac{\mathbf{d}^2F(\rho)}{\mathbf{d}\rho^2}+[(2l+2\kappa+2)\rho+(2l+d-2)\rho^{-1}]\frac{\mathbf{d}F(\rho)}{\mathbf{d}\rho}+[l(l+1)+\kappa(2l+1)+\kappa^2]F(\rho)=0.
\ee Thus, we have \be F(\rho)=\ _2F_1\left(\frac{l+\kappa+1}{2},
\frac{l+\kappa}{2}, l+\frac{d-1}{2};-\rho^2\right). \ee
Consequently, we get solution for Eq. (\ref{time}), (\ref{radi})
respectively, \be
T(\tau)=(\cosh\tau)^{-\frac{1}{2}(\frac{d}{2}-1)}\cdot\left\{
                   \begin{array}{l}
                   P_{\nu}^{\mu}(-i\sinh\tau) \\
                   \newline              \\
                   Q_{\nu}^{\mu}(-i\sinh\tau)
                   \end{array}\right.  ~,    \\
                   \newline \nonumber    \\
\label{Urho}U(\rho)=\rho^l(1+\rho^2)^{\frac{\kappa}{2}}\
_2F_1\left(\frac{l+\kappa+1}{2}, \frac{l+\kappa}{2},
l+\frac{d-1}{2};-\rho^2\right). \ee

A quantum field theory should be $CPT$ invariant. To the end, we set
\be \mu=0~~{\rm and}~~ \nu\in\mathbb{R}. \ee It means that
 \be \frac{(d-2)^2}{4}=\epsilon~~{\rm and}~~\kappa=\frac{d-2}{2}. \ee
 By making use of the relation between the associate Legendre function
and hypergeometric fuction\cite{21} \be\label{first Legendre}
P_{\nu}^{\mu}(z) &=& \frac{2^{\nu}\Gamma(\nu+\frac{1}{2})z^{\nu+\mu}
                      (z^2-1)^{-\mu/2}}{\Gamma(\frac{1}{2})\Gamma(1+\nu-\mu)}\ _2F_1\left(\frac{1-\nu-\mu}{2}, -\frac{\nu+\mu}{2},
                      \frac{1}{2}-\nu; z^{-2} \right)                         \nonumber\\
                    & &+\frac{2^{-\nu-1}\Gamma(-\nu-\frac{1}{2})z^{-\nu+\mu-1}
                      (z^2-1)^{-\mu/2}}{\Gamma(\frac{1}{2})\Gamma(-\nu-\mu)}\ _2F_1\left(\frac{2+\nu-\mu}{2}, \frac{1+\nu-\mu}{2},
                      \frac{3}{2}+\nu; z^{-2} \right) ~~,\\
\label{second Legendre}Q_{\nu}^{\mu}(z) &=& \frac{e^{\mu\pi
i}}{2^{\nu+1}}\frac{\Gamma(\nu+\mu+1)
                       \Gamma(\frac{1}{2})}{\Gamma(\nu+\frac{3}{2})}(z^2-1)^{\mu/2}
                       z^{-\nu-\mu-1}\nonumber\\
                       &&~~~~~~~\times\ _2F_1\left( \frac{\nu+\mu+1}{2}, \frac{\nu+\mu+2}{2},
                       \nu+\frac{3}{2}; z^{-2} \right) ~,
                      \ee
we can get
 \be\label{antipodal T}
T(-\tau)&=&T^\ast(\tau), \\ \label{antipodal
U}U(-\rho)&=&(-1)^lU^\ast(\rho). \ee
 From the relations (\ref{antipodal Ylm}), (\ref{antipodal T}) and (\ref{antipodal U})
we get \be \label{mode1}\phi(-x)=\phi^\ast(x). \ee
 In Sec. 5, we will show that property (\ref{mode1}) guarantees
that the scalar field theory is $CPT$ invariant.

\subsection{Spinor field  }
The Dirac equation in curved spacetime has been discussed
extensively by Brill and Wheeler\cite{17}, and Kibble\cite{18}. With
the help of the so-called tetrad or vierbein formalism\cite{19}, one
can convert general tensors into local, Lorentz-transforming tensors
and shift the additional spacetime dependence into vierbeins. The
metric of curved space $g_{ij}$ is related to the metric of
Minkowski space $\eta_{ab}$ by\be g_{ij}=e_i^a e_j^b \eta_{ab}, \ee
where $e_i^a$ is the so called vierbein. Converting the derivative
$\partial_i$ into the covariant derivative $\nabla_i$, one yields
the covariant Dirac equation in curved space\be
(i\gamma^i\nabla_i+im)\Psi=0,  \ee where \be
\nabla_i=\partial_i+\Gamma_i,  \ee $\gamma^i=e_a^i\gamma^a$ are the
curved space counterparts of the Dirac $\gamma$ matrices which
satisfy\be \{\gamma^i, \gamma^j\}=2g^{ij}, \ee and the connection is
given as \be \Gamma_i=\frac{1}{2}\Sigma^{ab}e_a^j(\partial_i
e_{bj}+\Gamma_{ij}^k e_{bk}).  \ee Here $\Sigma^{ab}$ is the
generator of the Lorentz group given as \be
\Sigma^{ab}=\frac{1}{4}[\gamma^a, \gamma^b],  \ee and
$\Gamma_{ij}^k$ is the Christoffel symbol defined as \be\label{Dirac
equation} \Gamma_{ij}^k=-\frac{1}{2}g^{kl}(\partial_i
g_{jl}+\partial_j g_{il}-\partial_l g_{ij}).  \ee The Dirac matrices
is defined as follows (remind that the spacetime dimension is even).
\\From the representations of Clifford algebra, we can give an
iterative expression of Dirac matrices starting from 2-dimensional
space, where \be \gamma^0={\begin{bmatrix} 0 & 1\\-1 & 0
\end{bmatrix}}, \ \gamma^1={\begin{bmatrix} 0 & 1\\1 & 0
\end{bmatrix}}.  \ee Then, in $d=2n+2$, we have \be
\gamma'^a=\gamma^a\bigotimes{\begin{bmatrix} -1 & 0\\0
& 1 \end{bmatrix}}, \ a=0, \cdots, d-3, \\
\gamma'^{d-2}=I\bigotimes{\begin{bmatrix} 0 & 1\\1 & 0
\end{bmatrix}}, \ \gamma'^{d-1}=I\bigotimes{\begin{bmatrix} 0 & -i\\i
& 0 \end{bmatrix}}, \ee with $\gamma^a$ the $2^n\times2^n$ Dirac
matrices in $(d-2)$ dimension and $I$ the $2^n\times2^n$ identity.

In the Beltrami coordinates with metric (\ref{second metric}), we
can write the Dirac equation (\ref{Dirac equation}) as
\be\label{Dirac equation1}
& &\Bigg\{i\cosh\tau\gamma_0\left(\partial_\tau+\frac{d-1}{2}\tanh\tau\right)-i\bigg[\gamma_1\left((1+\rho^2)\partial_\rho+\frac{d-2}{2}\rho^{-1}\right)\nonumber\\
& &+\sqrt{(1+\rho^2)}\rho^{-1}\Big[\gamma_2\left(\partial_{\theta_1}+\frac{d-3}{2}{\rm ctg}\theta_1\right)+\sin^{-1}\theta_1\gamma_3\left(\partial_{\theta_2}+\frac{d-4}{2}{\rm ctg}\theta_2\right)\nonumber\\
&
&+\cdots+\sin^{-1}\theta_1\cdots\sin^{-1}\theta_{d-3}\gamma_{d-1}\partial_{\theta_{d-2}}\Big]\bigg]-im\cosh\tau\Bigg\}\Psi(\tau,
\rho, \theta_1, \cdots, \theta_{d-2})=0. \ee By rewriting $\Psi$ as
\be \Psi(\tau, \rho, \theta_1, \cdots,
\theta_{d-2})=\cosh^{-\frac{d-1}{2}}\tau\sin^{-\frac{d-3}{2}}\theta_1\cdots\sin^{-\frac{1}{2}}\theta_{d-3}\psi(\tau,
\rho, \theta_1, \cdots, \theta_{d-2}), \ee we transform the equation
(\ref{Dirac equation1}) as \be\label{Dirac equation2}
\cosh\tau(\partial_\tau+m\gamma_0)\psi&=&\bigg[-\gamma_0\gamma_1\left((1+\rho^2)\partial_\rho+\frac{d-2}{2}\rho^{-1}\right)\nonumber\\
& &-i\sqrt{(1+\rho^2)}\rho^{-1}\gamma_1\hat{K}(\theta_1, \cdots,
\theta_{d-2})\bigg]\psi,  \ee where the Hermitian operator $\hat{K}$
is introduced as
 \be
\hat{K}=i\gamma_0\gamma_1(\gamma_2\partial_{\theta_1}
+\sin^{-1}\theta_1\gamma_3\partial_{\theta_2}+\cdots
+\sin^{-1}\theta_1\cdots\sin^{-1}\theta_{d-3}\gamma_{d-1}\partial_{\theta_{d-2}}).
\ee
 Analogous to the discussing in paper\cite{c}\cite{20}, we can
conclude that $\hat{K}$ commutes with $\hat{h}$\be
\hat{h}=-\gamma_0\gamma_1\left((1+\rho^2)\partial_\rho
+\frac{d-2}{2}\rho^{-1}\right)-i\sqrt{(1+\rho^2)}\rho^{-1}\gamma_1\hat{K},
\ee where $\hat{h}$ is related to the total angular momentum. Since
$\hat{K}$ commutes with $\hat{h}$, they have common eigenfunction
$\omega$ \be \hat{K}\omega&=&k\omega, \ k=0, \pm1, \pm2, \cdots, \\
-\hat{h}\omega&=&E\omega, \  E\in\mathbb{R}. \ee Then equation
(\ref{Dirac equation2}) can be transformed as\be
\left[\cosh\tau(\partial_\tau+m\gamma_0)+E\right]\omega=0,\\
\left[-\gamma_0\gamma_1\left((1+\rho^2)\partial_\rho
+\frac{d-2}{2}\rho^{-1}\right)-i\sqrt{(1+\rho^2)}\rho^{-1}\gamma_1
k+E\right]\omega=0. \ee By making use of similitude transformation
of Dirac matrices, one can always transform $\gamma_0$ and
$\gamma_1$ as\be \gamma_0=i{\begin{bmatrix} I & 0\\0 & -I
\end{bmatrix}}, \ \gamma_1={\begin{bmatrix} 0 & B\\B & 0
\end{bmatrix}}, \ee where $I$ is the $2^{d-2}\times2^{d-2}$ identity
and $B$ the $2^{d-2}\times2^{d-2}$ matrix satisfying $B^2=I$. The
transformation make the solving of equation (\ref{Dirac equation2})
easier.

For the sake of convenience, we write the eigenfunction $\omega$
into two-component spinor function form \be
\omega=D_k(\theta_1, \cdots, \theta_{d-2}){\begin{bmatrix} \varphi\\
\chi\end{bmatrix}}, \ee where $D_k(\theta_1, \cdots, \theta_{d-2})$
is the eigenfunction of operator $\hat{K}$. In the two-component
spinor form, the equation of field is\be\label{eq1}
\left[\cosh\tau(\partial_\tau+im)+E\right]\varphi=0, \ee
\be\label{eq2}
\left[(1+\rho^2)\partial_\rho+(1+k\sqrt{1+\rho^2})\right]
\left[(1+\rho^2)\partial_\rho+(1-k\sqrt{1+\rho^2})\right]\varphi+E^2\varphi=0.
\ee \be\label{eq3} \left[\cosh\tau(\partial_\tau-im)+E\right]\chi=0,
\ee \be\label{eq4}
\left[(1+\rho^2)\partial_\rho+(1-k\sqrt{1+\rho^2})\right]
\left[(1+\rho^2)\partial_\rho+(1+k\sqrt{1+\rho^2})\right]\chi+E^2\chi=0.
\ee The difference between the equations (\ref{eq1}), (\ref{eq2})
and the equations (\ref{eq3}), (\ref{eq4}) is just changing $m$ to
$-m$ and $k$ to $-k$. So that, we just need solve equations
(\ref{eq1}) and (\ref{eq2}).

The solution of equation (\ref{eq1}) is \be
\varphi(\tau)=e^{-im\tau}(\tanh\tau+{\rm sech}\tau)^{-Ea}. \ee The
solution of equation (\ref{eq2}) is more complicate. To solve it, we
make a transformation, \be
y=\frac{1-\sqrt{1+\rho^2}}{1+\sqrt{1+\rho^2}}. \ee Then, the
equation (\ref{eq2}) is transformed into the form \be
& &y(1-y)\frac{\mathbf{d}^2\varphi(y)}{\mathbf{d}y^2}+\frac{(d-5)y+(d-1)}{2}\frac{\mathbf{d}\varphi(y)}{\mathbf{d}y}+\frac{1}{4y(1-y)}\nonumber\\
&
&\times\left[k(1-y^2)-k^2(1-y)^2+2(d-2)y-4E^2y+\frac{(d-2)(d-4)}{4}(1+y)^2\right]\varphi(y)=0.
\ee By choosing \be \varphi(y)=y^\nu(1-y)^\mu F(y),\ ~
\nu=\pm\left(\frac{k}{2}-\frac{1}{4}\right)-\frac{d-3}{4}, \ ~
\mu=\pm E+\frac{d-2}{2}, \ee we get \be
y(1-y)\frac{\mathbf{d}^2F(y)}{\mathbf{d}y^2}+\left[\left(2\nu+\frac{d-1}{2}\right)-\left(2\nu+2\mu-\frac{d-5}{2}\right)y\right]\frac{\mathbf{d}F(y)}{\mathbf{d}y}\nonumber\\
-\left[\nu\left(\nu-\frac{d-3}{2}\right)+2\nu\mu+\mu\left(\mu-\frac{d-3}{2}\right)-\frac{k(k+1)}{4}+\frac{(d-2)(d-4)}{16}\right]F(y)=0.
\ee The solution of the above equation is of the form \be
F(y)=\left\{
                   \begin{array}{l}
                   \ _2F_1(k+\frac{1}{2}\pm E, \pm E,k+\frac{1}{2};y) \\
                   \newline              \\
                   \ _2F_1(1\pm E, \frac{1}{2}-k\pm E,\frac{3}{2}-k;y)
                   \end{array}\right.  ~. \ee
 Then, the solution
$\varphi(\tau, \rho)$ is \be \varphi(\tau,
\rho)=\varphi(\tau)\left(\frac{1-\sqrt{1+\rho^2}}{1+\sqrt{1+\rho^2}}\right)^\nu\left(\frac{2\sqrt{1+\rho^2}}{1+\sqrt{1+\rho^2}}\right)^\mu
F\left(\frac{1-\sqrt{1+\rho^2}}{1+\sqrt{1+\rho^2}}\right). \ee

In the final, we get the solution of the Dirac spinor function
$\Psi$ in Beltrami coordinates \be
\Psi=\cosh^{-\frac{d-1}{2}}\tau\sin^{-\frac{d-3}{2}}\theta_1\cdots\sin^{-\frac{1}{2}}\theta_{d-3}D_k(\theta_1, \cdots, \theta_{d-2}){\begin{bmatrix}\varphi(\tau, \rho)\\
\chi(\tau, \rho)\end{bmatrix}}. \ee

\section{Field quantization and CPT invariance}

Consider the symmetric two-point function \be G^{(1)}_{\la}(x,
y)=\langle\la|\Phi(x)\Phi(y)+\Phi(y)\Phi(x)|\la\rangle \ee in a de
Sitter-invariant state $|\la\rangle$. We assume $|\la\rangle$ is
invariant under the full disconnected group $O(1, 4)$, so that
$G^{(1)}_{\la}(x, y)$ can only depend upon the geodesic distance
$d(x, y)$. Because the geodesic distance $d(x, y)$ is a function of
$Z(x, y)$, the two-point function obeys the scalar field equation
\be (\Box-m^2)G^{(1)}(x, y)=0, \ee which can be written in terms of
$Z$, as \be \label{1}
\left[(Z^2-1)\frac{\mathbf{d}^2}{\mathbf{d}Z^2}+d\cdot
Z\frac{\mathbf{d}}{\mathbf{d}Z}+m^2\right]G^{(1)}(Z)=0. \ee This
wave equation is invariant under $Z\to-Z$. Thus the solution of
(\ref{1}) is of the form \be \label{2} G^{(1)}(Z)&=&c_1f(Z)+c_2f(Z),
\\f(Z)&=&_2F_1\left(h_+, h_-, \frac{d}{2}; \frac{1+Z}{2}\right), \\h_+h_-&=&m^2,  \ee
where $\ _2F_1\left(h_+, h_-, \frac{d}{2}; \frac{1+Z}{2}\right)$ is
the hypergeometric function, $c_1$ and $c_2$ are constants.

The general solution (\ref{2}) has two poles at $Z=1$ and $Z=-1$. If
$x$ is on the light cone of $y$, one has $Z(x,y)=1$. This is
analogous to the short-distance singularity along the light cone of
Minkowski space. Similarly, if $x$ is on the light cone of
$\bar{y}$, then $Z=-1$. We see that, singularity in de Sitter space
comes in pairs.

A real scaler field $\Phi_s$ on the elliptic de Sitter space, is
expressed in terms of a scalar field $\Phi$ on ordinary de Sitter
space,  as \be \Phi_s(x)=\frac{1}{\sqrt{2}}[\Phi(x)+\Phi(\bar x)]~.
\ee It is symmetric under action of the antipodal operator $A$.
\\Thus,  the Green function on the elliptic de Sitter space is given as
\be G^{(1)}_{s \la}(x,
y)=\langle\la|\Phi_s(x)\Phi_s(y)+\Phi_s(y)\Phi_s(x)|\la\rangle. \ee
 In terms of $G^{(1)}_{\la}(x, y)$, $G^{(1)}_{s \la}(x,
y)$ can be expressed as \be G^{(1)}_{s \la}(x,
y)=\frac{1}{2}[G^{(1)}_{\la}(x, y)+G^{(1)}_{\la}(x, \bar
y)+G^{(1)}_{\la}(\bar x, y)+G^{(1)}_{\la}(\bar x, \bar y)]. \ee
Noticing that in de Sitter space $G^{(1)}_{\la}(x, y)$ is a function
of $Z(x, y)$,  and making use of Equation(\ref{Z}), we obtain the
following relations, \be G^{(1)}_{\la}(x, y)&=&G^{(1)}_{\la}(\bar x,
\bar y),
\\G^{(1)}_{\la}(\bar x, y)&=&G^{(1)}_{\la}(x, \bar y). \ee
Therefore, we have \be \label{sym green} G^{(1)}_{s \la}(x,
y)=G^{(1)}_{\la}(x, y)+G^{(1)}_{\la}(x, \bar y). \ee

Before considering quantization of fields on the elliptic de Sitter
space, let us recall briefly the canonical quantization of a scalar
field in de Sitter space.

Normally, one expands the scalar field in terms of its modes as \be
\label{mode exp}
\Phi(x)=\sum_n~[a_n\phi_n(x)+a_n^\dag\phi_n^\ast(x)].  \ee The modes
$\phi_n(x)$ satisfy the wave equation \be \label{wave
equation}(\Box-m^2)\phi_n(x)=0. \ee They are orthonormal each other
with the Klein-Gordon inner product \be (\phi_n,
\phi_m)=-i\int_\Sigma d\Sigma^\mu(\phi_n\partial_\mu
\phi_m^\ast-\phi_m\partial_\mu \phi_n^\ast),  \ee where $\Sigma$ is
a Cauchy surface. The operators $a_n$ and $a_n^\dag$ satisfy
commutation relations \be [a_n, a_m^\dag]=\delta_{nm}. \ee And the
vacuum state $|\Omega\rangle$ is defined uniquely as \be
a_n|\Omega\rangle=0. \ee The Wightman function is defined by \be
G_\Omega(x,
y)\equiv\langle\Omega|\Phi(x)\Phi(y)|\Omega\rangle=\sum_n\phi_n(x)\phi_n^\ast(y).
\ee There is a unique state, the ``Euclidean" vacuum $|E\rangle$.
The short distance singularities in $G_E^{(1)}$ have the Hadamard
form, the coefficient of the $\sigma^{-1}=2/d(x, y)^{-2}$ is the
same as in flat space. The Wightman function in state $|E\rangle$ is
\be G_E(x, y)=\langle
E|\phi(x)\phi(y)|E\rangle=\frac{\Gamma(h_+)\Gamma(h_-)}{(4\pi)^d/2\Gamma(\frac{d}{2})}\
_2F_1\left(h_+, h_-, \frac{d}{2};\frac{1+Z(x, y)}{2}\right).  \ee In
de sitter space, there is a family of de Sitter invariant Green
functions parametrized by $\alpha$\cite{a1}--\cite{a6}. The
``Euclidean" vacuum corresponds to $\alpha=0$. The Green function
for these vacuum can be expressed as \be G_\alpha^{(1)}(x,
y)=\cosh2\alpha~ G_0^{(1)}(x, y)+\sinh2\alpha~ G_0^{(1)}(x, \bar y).
\ee In the elliptic de Sitter space, by making use of Eq.(\ref{sym
green}), we have \be G_{s \alpha}^{(1)}=e^{2\alpha}[G_E^{(1)}(x,
y)+G_E^{(1)}(x, \bar y)]. \ee Unlike the case of de Sitter space, in
the elliptic de Sitter space even for $\alpha=0$, $G_{s
\alpha}^{(1)}$ contains singularities both for $Z(x, y)=1$ and $Z(x,
y)=-1$ and with the same strength. However, we regard the $\alpha$
vacua for $\alpha\neq0$ as unphysical. Since their Green functions
do not have the same short distance singularities as in flat space.
As discussed in \cite{12}, in standard renormalization theory the
subtractions include nonlocal contributions to the effective action
for $\alpha\neq0$. I

Going around from a point in de Sitter space to its antipodal point
has the effect of acting on tangent space by $PT$ and the
$\mathbb{Z}_2$ map also requires charge conjugation $C$.  The
cumulative effect is to relate a point of its antipodal point by
$CPT$\cite{10}. After the $\mathbb{Z}_2$ identification,  the effect
of $CPT$ is just relate $\Phi(x)$ to itself in the elliptic de
Sitter space. The mode expansion of complex scalar field $\Phi$ is
\be \Phi(x)=\sum_n~[a_n\phi_n(x)+b_n^\dag\phi^\ast_n(x)].
 \ee
We define the action of $\mathcal{P}\mathcal{T}$ by \be
\mathcal{P}\mathcal{T}\Phi(x)\mathcal{T}^{-1}\mathcal{P}^{-1}=\Phi(-x),
\ee and the action of $\mathcal{C}$ by \be
\mathcal{C}\Phi(x)\mathcal{C}^{-1}=\Phi^\dag(x).  \ee Then, we
get\be
\mathcal{P}\mathcal{T}a_n\mathcal{T}^{-1}\mathcal{P}^{-1}=b_n^\dag,~~and~~~
\mathcal{P}\mathcal{T}b_n^\dag\mathcal{T}^{-1}\mathcal{P}^{-1}=a_n,\ee
\be \mathcal{C}a_n\mathcal{C}^{-1}=b_n,~~and~~~
\mathcal{C}b_n\mathcal{C}^{-1}=a_n. \ee This is in agreement with
analytic proof of the CPT theorem on quantum field theory in de
Sitter space\cite{25}-\cite{28}. In fact, one can check without
difficulty that the quantum field theory in the Beltrami coordinate
of elliptic de Sitter space satisfies the so called assumptions of
covariance, weak spectral condition and locality.

In the limit of the Hubble constant $\frac{1}{a}\rightarrow0$, the
de Sitter space simply reduces to two copies of Minkowski space. The
second copy is the $CPT$ conjugate of the first. And the elliptic de
Sitter space goes to Minkowski space.

\section{Conclusions and remarks}
In this paper, we have discussed consequences of Schr\"{o}dinger's
antipodal identification of de Sitter space on quantum field theory.
The elliptic $\mathbb{Z}_2$ identification provides observers with
complete information.  Unfortunately, a general $d$-dimensional
elliptic de Sitter space $dS_d/\mathbb{Z}_2$ is non-orientable. This
is a fatal problem for constructing a quantum field theory. We
calculated the first Stiefel-Whitney class of the elliptic de Sitter
space and showed that the manifold $dS_d/\mathbb{Z}_2$ is orientable
for odd $d$. Furthermore, we have noted that the second
Stiefel-Whitney class $\omega_2(dS_d/\mathbb{Z}_2)$ vanished while
the dimension of the elliptic de Sitter space equal to $3~{\rm
mod}~4$. Thus, for $3~{\rm mod}~4$-dimensional elliptic de Sitter
space, there exists globally defined spinors.  In Beltrami
coordinates, we have given exact solutions of scalar field and
spinor field. The CPT invariance of the quantum field theory on the
elliptic de Sitter space was shown explicitly.

Though the asymptotic geometry of elliptic de Sitter space consists
of only a single boundary $S^{d-1}$, the $\mathbb{Z}_2$
identification makes the boundary of elliptic de Sitter space much
more complicate. The discussing about $S$ matrix, and interaction
between scalar field and spinor field will be published separately.

 The striking elliptic interpretation given by Schr\"{o}dinger is
the implementation of observer complementarity.  However,
Schr\"{o}dinger did not tell us how this can be done in detail. Here
we suggest that the implementation of observer complementarity is
relate to the entanglement entropy. According to the big bang model
of universe observers are close enough. We simply considered the
total quantum system as a pure state. Next, we divide the total
system into two subsystems and they represent the observer and his
antipodal observer respectively. The universe will eventually become
an asymptotic de Sitter Space. Then, knowing the entanglement
entropy of one system means knowing the entanglement entropy of
antipodal system.  This can be regard as the implementation of
observer complementarity.

\vspace{1.5cm}

\centerline{\bf \large Acknowledgement}

\vspace{0.3cm}

We would like to thank Prof. H. Y. Guo and C.-G. Huang for helpful
discussion.  We are indebted to Prof. J.-Z. Pan who sent us his
useful unpublished notes on orientation of elliptic de Sitter space.
The work was supported by the NSF of China under Grant Nos. 10375072
and 10575106.
\begin{appendix}
\section{Vierbein and Chirstoffel symbol}

 In this appendix, we present the detail
vierbeins and Chirstoffel symbol used in the Sec. 4.3.

Vierbein:\be e_i^a={\rm
diag}(1,\cosh\tau(1+\rho^2)^{-1},\cosh\tau\cdot\rho(1+\rho^2)^{-\frac{1}{2}}(1,\sin\theta_1,\cdots,\sin\theta_1\cdots\sin\theta_{d-2})).
 \ee
 By making use of the substitution
 \be R=\cosh t \ {\rm and}\ \lambda=1+\rho^2~,\ee
 we may express the Chirstoffel symbol as:\be
 \Gamma^\tau_{\rho\rho}=-R\frac{\mathbf{d}R}{\mathbf{d}\tau}\lambda^{-2},~~~~
 \Gamma^\rho_{\rho\rho}=2\lambda^{-1}\rho,~~~~
 \Gamma^\rho_{\tau\rho}=-R^{-1}\frac{\mathbf{d}R}{\mathbf{d}\tau},~\\
 \Gamma^\tau_{\theta_1\theta_1}=-R\frac{\mathbf{d}R}{\mathbf{d}\tau}\lambda^{-1}\rho^2,~~~~
 \Gamma^\tau_{\theta_n\theta_n}=\sin^2\theta_{n-1}\Gamma^\tau_{\theta_{n-1}\theta_{n-1}}~(n=2,\cdots,d-2),\\
 \Gamma^\rho_{\theta_1\theta_1}=\rho,~~~~
 \Gamma^\rho_{\theta_n\theta_n}=\sin^2\theta_{n-1}\Gamma^\rho_{\theta_{n-1}\theta_{n-1}}~(n=2,\cdots,d-2),\\
 \Gamma^{\theta_{n-1}}_{\tau\theta_{n-1}}=-R^{-1}\frac{\mathbf{d}R}{\mathbf{d}\tau}~(n=2,\cdots,d-2),\\
 \Gamma^{\theta_{n-1}}_{\rho\theta_{n-1}}=-\rho^{-1}\lambda^{-1}~(n=2,\cdots,d-2),\\
 \Gamma^{\theta_{n-1}}_{\theta_n\theta_n}=\sin\theta_{n-1}\cos\theta_{n-1}~(n=2,\cdots,d-2),\\
 \Gamma^{\theta_n}_{\theta_m\theta_n}=-{\rm ctg}\theta_m~~[(n=2,\cdots,d-2)~{\rm and}~(m=1,\cdots,d-3)~m<n],\\
 \Gamma^{\theta_m}_{\theta_{l+1}\theta_{l+1}}=\sin^2\theta_l
 \Gamma^{\theta_m}_{\theta_l\theta_l}~~[(l=2,\cdots,d-3)~{\rm and}~(m=1,\cdots,d-3)].
 \ee
\end{appendix}

%
%
\end{document}